\documentclass[aps,longbibliography,showpacs,twocolumn,superscriptaddress,amsmath,amssymb,verbatim]{revtex4-2}
\usepackage[thinlines]{easytable}
\usepackage{graphicx}
\usepackage{lmodern}
\usepackage{epstopdf}
\usepackage{dcolumn}
\usepackage{bm}
\usepackage{subfigure}
\usepackage{makecell}
\usepackage{amsmath} 
\usepackage[pdftex,colorlinks=true,citecolor=blue,linkcolor=blue,urlcolor=blue,bookmarks=true]{hyperref}

\usepackage[utf8]{inputenc}

\usepackage{xcolor}
\usepackage{tikz}
\usepackage[version=4]{mhchem}

\usepackage{makecell}

\usepackage{color}
\usepackage{textcomp}
\definecolor{red}{rgb}{1,0,0}

\definecolor{blue}{rgb}{0,0,1}

\definecolor{green}{rgb}{0,1,0}

\begin{document}
	\preprint{APS}

\title{Multistage development of short-range spin correlations and weak magnetic order in the two coupled trillium lattices of \ce{K2Fe2(MoO4)(PO4)2}}
\author{J. Khatua}
\thanks{These authors contributed equally to this work.}
\affiliation{Department of Physics, Sungkyunkwan University, Suwon 16419, Republic of Korea}
\author{Sritharan Krishnamoorthi}
\thanks{These authors contributed equally to this work.}
\affiliation{Institute of Physics, Academia Sinica, Taipei 11529, Taiwan}
\author{Changhyun Koo}
\affiliation{Department of Physics, Sungkyunkwan University, Suwon 16419, Republic of Korea}
\author{Gyungbin Ban}
\affiliation{Department of Physics, Sungkyunkwan University, Suwon 16419, Republic of Korea}
\author{Taeyun Kim}
\affiliation{Department of Physics, Sungkyunkwan University, Suwon 16419, Republic of Korea}
\author{Yugo Oshima}
\affiliation{RIKEN Pioneering Research Institute, Wako, Saitama 351-0198, Japan}
\author{Marc Uhlarz}
\affiliation{Dresden High Magnetic Field Laboratory (HLD-EMFL), Helmholtz-Zentrum Dresden-Rossendorf, 01328 Dresden, Germany}
\author{John Wilkinson}
\affiliation{ISIS Pulsed Neutron and Muon Source, STFC Rutherford Appleton Laboratory,
	Harwell Campus, Didcot, Oxfordshire OX110QX, UK}
\author{Peter J. Baker}
\affiliation{ISIS Pulsed Neutron and Muon Source, STFC Rutherford Appleton Laboratory,
	Harwell Campus, Didcot, Oxfordshire OX110QX, UK}
\author{Kyeong Jun Lee}
\affiliation{Department of Physics, Chung-Ang University, Seoul 60974, Republic of Korea}
\author{Mani Shankar}
\affiliation{Institute of Physics, Academia Sinica, Taipei 11529, Taiwan}
\author{Seo Hyoung Chang}
\affiliation{Department of Physics, Chung-Ang University, Seoul 60974, Republic of Korea}
\author{R. Sankar}
\email[]{sankarndf@gmail.com}
\affiliation{Institute of Physics, Academia Sinica, Taipei 11529, Taiwan}
\author{Kwang-Yong Choi}
\email[]{choisky99@skku.edu}
\affiliation{Department of Physics, Sungkyunkwan University, Suwon 16419, Republic of Korea}

\date{\today}

\begin{abstract}
Trillium lattices, where magnetic ions form a chiral network of corner-sharing triangles, offer a three-dimensional magnetic framework that can host fragile classical spin-liquid states. Herein, we report on the magnetization, specific heat, electron spin resonance (ESR), and muon spin relaxation ($\mu$SR) of  K$_{2}$Fe$_{2}$(MoO$_{4}$)(PO$_{4}$)$_{2}$ single crystals. Magnetization measurements reveal strong antiferromagnetic interactions coexisting with weak magnetic order at $T_{\rm N} = 5.2$~K, as evidenced by a $\lambda$-like anomaly observed in the magnetic susceptibility, a critical enhancement of the muon spin relaxation rate and the wipeout of the ESR signal as the temperature approaches $T_{\rm N}$. Above $T_{\rm N}$, two distinct developments of short-range spin correlations are identified at $T_{\rm H} = 34$~K and $T_{\rm L} = 10$~K, supported by magnetic specific heat anomalies and the temperature dependence of the ESR linewidth and $g$-factor.
 Upon cooling below $T_{\rm N}$, an anomaly appears at $T^{*} = 3.2$~K in thermodynamic observables and the muon spin relaxation rate, indicative of spin reorientation driven by residual interactions. Despite the presence of magnetic order, $\mu$SR experiments reveal dynamically fluctuating spins persisting even in the ordered state. Moreover, the suppression of $T_{\rm N}$ under applied magnetic fields ($\mu_{0}H \geq 2$~T) suggests that K$_{2}$Fe$_{2}$(MoO$_{4}$)(PO$_{4}$)$_{2}$ constitutes  a promising candidate for exploring field-induced spin-liquid behavior in three-dimensionally coupled trillium lattices.

\end{abstract}
\maketitle

\section{Introduction}
In recent years, the quest for many-body quantum phenomena that give rise to exotic ground states in geometrically frustrated magnets has attracted considerable interest~\cite{KHATUA20231}. Among the most sought-after states is the quantum spin liquid (QSL), an elusive state of matter  characterized by the absence of magnetic order even at $T \to 0$ K, long-range quantum entanglement, and fractionalized excitations such as spinons~\cite{Balents2010}. Within the framework of resonating valence bond (RVB) theory~\cite{ANDERSON1973153}, two-dimensional (2D) frustrated magnets based on triangular and kagome lattices have long served as prime platforms for realizing QSL states~\cite{Savary_2016,scienceaay0668}. However, the extension of RVB physics to three dimensions remains a fundamental challenge~\cite{Glittum2025}.\\
 In this situation, the trillium lattice (space group $P2_{1}3$), a 3D chiral network of corner-sharing equilateral triangles, has recently emerged as a promising geometry for frustration-driven phenomena~\cite{PhysRevB.74.224441,Khatua_2025,PhysRevLett.128.177201,PhysRevLett.127.157204}. Indeed, theoretical studies predict a rich variety of quantum phases on the trillium lattice, including (i) two $Z_{2}$  and one gapless $U$(1) QSLs within the projective symmetry group framework \cite{3nmp-1vt2}, (ii) a fractal spin liquid in the Newman–Moore model~\cite{PhysRevB.105.224410}, and (iii) a fragile topological classical spin liquid (CSL) in the classical Heisenberg model~\cite{PhysRevB.110.L020402,PhysRevB.109.174421,PhysRevB.111.134413}. In contrast to the well-studied $U$(1) CSL characterized by pinch-point correlations~\cite{annurev}, the fragile CSL exhibits an extensive ground-state degeneracy, an absence of long-range order, and exponentially decaying short-range spin correlations~\cite{w2m6-bs9h,PhysRevB.111.134413}.
\\ While intermetallic compounds with single trillium lattices of magnetic ions host multi-$Q$ states~\cite{PhysRevB.109.174437,PhysRevB.107.174408}, chiral phenomena~\cite{science1166767,13yj-sbgc}, and signatures of frustrated magnetism~\cite{PhysRevB.74.224441,PhysRevB.104.045145}, experimental exploration  of the theoretically proposed frustrated magnetism in insulating systems has been limited by the scarcity of suitable compounds. Recently, the search for exotic ground states in frustrated trillium compounds has gained renewed interest with the discovery of the langbeinite family K$_{2}$$M_{2}$(SO$_{4}$)$_{3}$ ($M$ =Ni$^{2+}$, Co$^{2+}$, etc.), which crystallizes in the noncentrosymmetric space group $P2_{1}3$ and consists of a double trillium lattice~\cite{PhysRevLett.127.157204,10.1063/5.0096942,PhysRevLett.131.146701,PhysRevB.109.184432,m8mj-slvl,Gonzalez2024}. Importantly, the inter-trillium exchange couplings in these systems generate a hypertrillium (or tetratrillium) magnetic network composed of three corner-sharing octahedra, leading to a highly frustrated spin model that has been proposed to support a fragile CSL ground state within the classical Heisenberg model~\cite{w2m6-bs9h}. \\ 
Besides low-spin candidates, Fe$^{3+}$ ($S=5/2$) based phosphate compounds in the langbeinite family have also become strong contenders for realizing fragile CSL  states. For instance, KSrFe$_2$(PO$_4$)$_3$, a double trillium compound with  K/Sr anti-site disorder, exhibits short-range spin freezing coexisting with persistent spin dynamics~\cite{k3lw-2567}. In contrast, K$_2$FeSn(PO$_4$)$_3$, a related compound with Fe/Sn magnetic site dilution, displays the coexistence of pronounced spin fluctuations and weak  magnetic order~\cite{lmsf-73hn}. These observations indicate that robust spin dynamics persist largely independent of the degree of structural disorder, pointing to spin-liquid-like correlations intrinsic to the hypertrillium lattice, while residual interactions lift the degeneracy only weakly, giving rise to fragile magnetic order. Although the precise microscopic origin of this behavior remains unclear, a key direction for future studies is to identify double trillium lattice candidates with minimal disorder, in which the hypertrillium network supports larger exchange energy scales. Such systems may stabilize spin-liquid features to higher temperatures, rendering them observable above the energy scale set by residual interactions.\\
In this context, we investigate the ground-state properties of 
K$_2$Fe$_2$(MoO$_4$)(PO$_4$)$_2$ single crystals, in which Fe$^{3+}$ ions 
form a double trillium lattice free from alkali-metal and magnetic-site disorder, with only a minor degree of exchange disorder possibly due to Mo/P site mixing, using magnetization, specific heat, electron spin resonance (ESR), and muon spin relaxation ($\mu$SR) 
measurements. Weak magnetic order is observed at 
$T_{\rm N}=5.2$~K in several probes, coexisting with strong 
antiferromagnetic correlations. Above $T_{\rm N}$, magnetization, 
specific heat, and ESR reveal two distinct regimes of short-range 
spin correlations at $T_{\rm H}=34$~K and $T_{\rm L}=10$~K. 
An additional anomaly at $T^{*}=3.2$~K indicates a spin reordering by residual interactions below $T_{\rm N}$. Notably, $\mu$SR uncovers 
persistent spin dynamics even in the ordered state, while the 
strong suppression of $T_{\rm N}$ under applied magnetic fields 
($\mu_{0}H \geq 2$~T) suggests that the weak magnetic order arises 
from subleading anisotropic or residual interactions in addition to dominant Heisenberg exchange interactions.

\section{EXPERIMENTAL DETAILS}
High-quality single crystals of potassium iron(III) molybdate phosphate, \ce{K2Fe2(MoO4)(PO4)2} (hereafter KFMPO), were grown using a high-temperature flux method. High-purity precursor materials, \ce{K2MoO4} (19.2~g), \ce{MoO3} (17.2~g), \ce{KPO3} (6.7~g), and \ce{Fe2O3} (4.8~g), were weighed in stoichiometric proportions and thoroughly ground in an agate mortar to ensure compositional homogeneity. The mixture was transferred into a platinum crucible and heated to $1000^{\circ}\mathrm{C}$ at a rate of $100^{\circ}\mathrm{C\,h^{-1}}$, followed by a dwell of 24~h to achieve complete melting and dissolution of the starting materials. The homogeneous melt was subsequently cooled from $1000^{\circ}\mathrm{C}$ to $500^{\circ}\mathrm{C}$ at a controlled rate of $2^{\circ}\mathrm{C\,h^{-1}}$, enabling the nucleation and growth of well-formed single crystals with high crystalline quality. After reaching $500^{\circ}\mathrm{C}$, the furnace was switched off and allowed to cool naturally to room temperature. Brown-colored single crystals suitable for structural and physical property measurements were recovered from the residual flux by leaching with hot deionized water.
\\
 Powder X-ray diffraction (XRD) measurements were carried out at room temperature on crushed single crystals of K$_{2}$Fe$_{2}$(MoO$_{4}$)(PO$_{4}$)$_{2}$ using a Bruker  D2 diffractometer with Cu K$_{\alpha}$ radiation ($\lambda = 1.54$~\AA). In addition, diffraction patterns for the (\textit{hhh}) planes were collected using high-resolution XRD (HR-XRD, Bruker AXS D8) equipped with a Cu K$_{\alpha}$ source.
\\ 
The \textit{dc} magnetic susceptibility and magnetization were measured using a superconducting quantum interference device vibrating sample magnetometer (SQUID-VSM, Quantum Design, USA) in the temperature range $2 \leq T \leq 300$~K and in applied magnetic fields of up to 9~T.
  On the other hand, the \textit{ac} magnetic susceptibility measurements were performed between 2 K to 20 K using the Quantum Design MPMS
SQUID system at three different frequencies. High-field magnetization data were recorded
at the Dresden High Magnetic Field Laboratory, sweeping a
magnetic field up to 55 T at 1.6 K using a nondestructive
pulsed magnet.\\\begin{figure*}
	\centering
	\includegraphics[width=\textwidth]{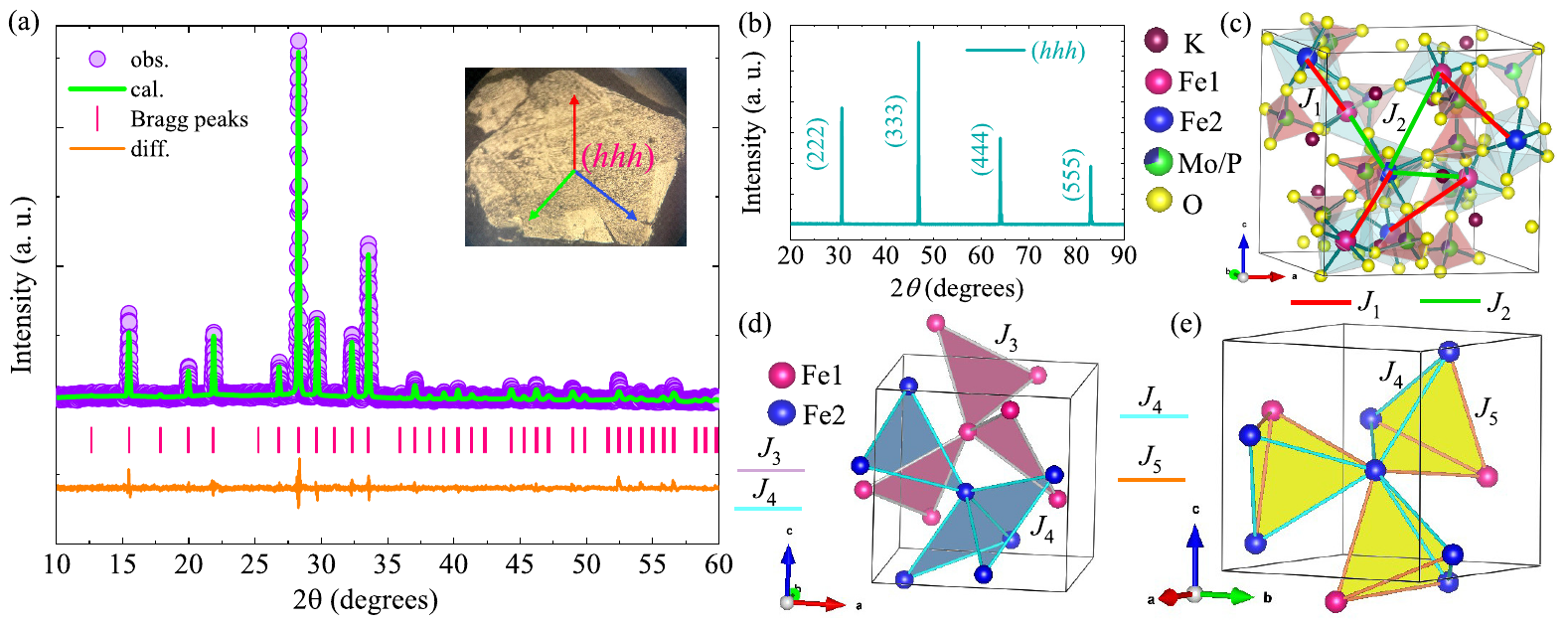}
	\caption{(a) Rietveld refinement of the powder \text{X}-ray diffraction pattern collected at room temperature. The colored curves correspond to the observed, calculated, and difference diffraction profiles, while the vertical pink bars indicate the Bragg reflection positions, as shown in the legend. The inset (top right) shows a photo of a  single crystal of \ce{K2Fe2(MoO4)(PO4)2}, with the top surface indicating to the ($hhh$) plane. (b) X-ray diffraction
		pattern of a single crystal of the (\textit{hhh}) plane.
		(c) Crystal structure of one unit cell of \ce{K2Fe2(MoO4)(PO4)2}, consisting of FeO$_6$ octahedra that share corners with Mo/P tetrahedra. 
		(d) Schematic illustration of the double trillium lattice formed by each crystallographic Fe$^{3+}$ site.
		(e) Hypertrillium lattice, where specific inter-trillium bond lengths  connect the Fe1 sites to the trillium networks of the Fe2 sites. }
	\label{XRD}
\end{figure*}Specific heat measurements were performed using a standard relaxation method with a physical property measurement
system (PPMS, Quantum Design, USA) in the temperature range 2 K $\leq$ $T$ $\leq$ 300 K and in magnetic fields up
to 9 T.  \\   
ESR measurements were conducted using a conventional X-band ($f$ = 9.12 GHz) ESR spectrometer (JEOL, JES-RE3X) in RIKEN. A single crystal of \ce{KFMPO} was loaded on a quartz rod, allowing the magnetic field to apply to [111] crystal orientation. A continuous $^4$He-flow cryostat enables controlling temperatures from 3.7 K to 280 K for the experiments.
\\ \\
$\mu$SR experiments were conducted
using the MuSR spectrometer at the ISIS Neutron
and Muon Source, STFC Rutherford Appleton Laboratory in the UK~\cite{ISISdatabase}. The initial muon spin polarization was aligned parallel to the beam direction. Multiple small single crystals of KFMPO were mounted on a silver sample holder using GE varnish and covered with a thin silver foil. The $\mu$SR spectra were acquired in the temperature range 4 K $\leq$ $T$ $\leq$ 60 K using a $^4$He cryostat, and at lower temperatures down to 0.11 K using a dilution refrigerator. All zero-field and longitudinal-field  $\mu$SR data were analyzed using the \text{W}i\text{MDA} software package~\cite{PRATT2000710}.
\begin{table}[b]
	\caption{\label{table} Refined structural parameters from Rietveld analysis of \text{X}-ray diffraction data at 300 K. (Space group: $P$2$_{1}$3, $a = b= c$ = 10.055 \AA , $\alpha = \beta = \gamma = 90^{\circ}$
		and $\chi^{2}$ = 1.39, \textit{R}$_{\rm wp}$ = 0.15, \textit{R}$_{\rm p}$ = 0.11)}
	\begin{tabular}{c c c c c  c c} 
		\hline \hline
		Atom & Wyckoff position & \textit{x} & \textit{y} &\textit{ z}& Occ.\\
		\hline 
		K${1}$ & 4$a$ & 0.708 \ \ & 0.708\ \ & 0.708 \  & 1 \\
		K${2}$ & 12$b$ & 0.932\ \ & 0.932\ \ & 0.932 \ \ & 1 \\
		Fe${1}$ & 12$b$ & 0.141 \ \ & 0.141 \ \ & 0.141 \ \ &1 \\
		Fe${2}$ & 12$b$ & 0.414 \ \ & 0.414 \ \ & 0.414 \ \ & 1 \\
		P & 12$b$ & 0.452 \ \ & 0.225 \ \ & 0.127 \ \ & 0.70 \\
		Mo & 12$b$ & 0.452\ \ &  0.227\ \ &  0.127 \ \ & 0.30 \\
		O${1}$ & 12$b$ & 0.305\ \ & 0.234\ \ & 0.077\ \ & 1 \\
		O${2}$ & 12$b$ & 0.546\ \ & 0.299\ \ & 0.019\ \ & 1 \\
		O${3}$ &  12$b$ & 0.503\ \ & 0.080\ \ & 0.152\ \ & 1 \\
		O${4}$ & 12$b$ & 0.474\ \ & 0.305\ \ & 0.260\ \ & 1 \\
		\\	
		\hline
	\end{tabular}
\end{table}
\begin{table*}[t]\label{EXCH}
	\centering
	\caption{Details of the magnetic exchange coupling paths in \ce{K2Fe2(MoO4)(PO4)2} with relevant bond angles. A close comparison with analogous exchange pathways in isostructural  compound \ce{KSrFe2(PO4)3} is also presented~\cite{k3lw-2567,10.1063/5.0096942}.}
	\label{tab:exchange}
	\begin{tabular}{ccc|ccc}
		\hline\hline
		\multicolumn{3}{c}{\text{\ce{K2Fe2(MoO4)(PO4)2}} ($a$ = 10.05 \AA, $\theta_{\rm CW}=-102$~K)}
		&
			\multicolumn{3}{c}{\text{\ce{KSrFe2(PO4)3}} ($a$ = 9.80 \AA, $\theta_{\rm CW}=-76$~K) \cite{k3lw-2567,10.1063/5.0096942}} \\
		\cline{1-3}\cline{4-6}
		\makecell[c]{Exchange\\interaction} & Bond length (\AA) & Exchange path
		& Bond length (\AA) &  Exchange path & \makecell[c]{Exchange\\interaction (K)} \\
		\hline
		 \makecell[c]{$J_1$, Dimer\\(Fe2--Fe1)}
		& 4.74 & \makecell[l]{Fe2--O4--P--O1--Fe1 \\
			$\angle$Fe2--O4--P = 154.1$^\circ$ \\
			$\angle$P--O1--Fe1 = 130.3$^\circ$}
		& 4.53 & \makecell[l]{Fe2--O3--P--O2--Fe1 \\
			$\angle$Fe2--O3--P = 144.8$^\circ$ \\
			$\angle$P--O2--Fe1 = 134.4$^\circ$} & $-1.5$ \\ \hline
		 \makecell[c]{$J_2$, Inter-Dimer\\(Fe1--Fe2)}
		& 5.03 & \makecell[l]{Fe1--O2--P--O3--Fe2 \\
			$\angle$Fe1--O2--P = 166.1$^\circ$ \\
			$\angle$P--O3--Fe2 = 150.5$^\circ$}
		& 4.92   & \makecell[l]{Fe1--O1--P--O4--Fe2 \\
			$\angle$Fe1--O1--P = 157.3$^\circ$ \\
			$\angle$P--O4--Fe2 = 148.6$^\circ$}  & $0$ \\ \hline
		 \makecell[c]{$J_3$, Trillium (Site1)\\(Fe1--Fe1)}
		& 6.16 & \makecell[l]{Fe1--O2--P--O1--Fe1 \\
			$\angle$Fe1--O2--P = 166.1$^\circ$ \\
			$\angle$P--O1--Fe1 = 130.3$^\circ$}
		& 6.02 & \makecell[l]{Fe1--O2--P--O1--Fe1 \\
			$\angle$Fe1--O2--P = 134.4$^\circ$ \\
			$\angle$P--O1--Fe1 = 157.3$^\circ$}  & $-1.6$ \\ \hline
		 \makecell[c]{$J_4$, Trillium (Site2)\\(Fe2--Fe2)}
		& 6.25 & \makecell[l]{Fe2--O3--P--O4--Fe2 \\
			$\angle$Fe2--O3--P = 150.4$^\circ$ \\
			$\angle$P--O4--Fe2 = 154.1$^\circ$}
		& 6.09  & \makecell[l]{Fe2--O3--P--O4--Fe2 \\
			$\angle$Fe2--O3--P = 144.8$^\circ$ \\
			$\angle$P--O4--Fe2 = 148.6$^\circ$} & $-3.9$ \\ \hline
		 \makecell[c]{$J_5$, Inter-Trillium\\(Fe2--Fe1)}
		& 6.06 & \makecell[l]{Fe2--O4--P--O2--Fe1 \\
			$\angle$Fe2--O4--P = 154.1$^\circ$ \\
			$\angle$P--O2--Fe1 = 166.1$^\circ$}
		& 5.95 & \makecell[l]{Fe2--O3--P--O1--Fe1 \\
			$\angle$Fe2--O3--P = 144.8$^\circ$ \\
			$\angle$P--O1--Fe1 = 157.3$^\circ$} & $-2.7$ \\ \hline
	\end{tabular}
\end{table*}
\section{RESULTS AND DISCUSSION}	
\subsection{X-ray diffraction and crystal structure}
To examine the phase purity and crystal structure, XRD measurements were performed at room temperature on crushed single crystals of \ce{KFMPO}. Rietveld refinement of the powder XRD data was carried out using the GSAS software, with the initial atomic coordinates taken from Ref.~\cite{Slobodyanik2012}. Figure~\ref{XRD}(a) presents the observed XRD pattern (maroon circles) together with the calculated profile from the Rietveld refinement (solid line) for \ce{KFMPO}. All reflections are well reproduced by the refinement and can be indexed within the cubic space group $P2_{1}3$. The absence of extra reflections confirms the high phase purity of the sample used in this study. The single-crystal XRD pattern corresponding to the (\textit{hhh}) plane is shown in Fig.~\ref{XRD}(b). Notably, crystallization in the noncentrosymmetric $P2_{1}3$ space group places the present compound in the B20 structural family, in which the lack of inversion symmetry intrinsically allows Dzyaloshinskii–Moriya (DM) interactions~\cite{,Khatua_2025,PhysRevB.70.075114}.\\
The refined coordinates and lattice parameters  are summarized in Table~\ref{table}. A representative unit cell of \ce{KFMPO} is shown in Fig.~\ref{XRD}(c). All constituent atomic sites are fully occupied, except for anti-site disorder between Mo$^{6+}$ and P$^{5+}$ ions. The  unit cell contains two crystallographically distinct Fe$^{3+}$ sites, each forming an FeO$_6$ octahedron with six nearest-neighbor oxygen ions. Each FeO$_6$ octahedron shares its six corners with six Mo/P tetrahedra. Conversely, each Mo/P tetrahedron is connected via its four corners to four Fe ions, enabling the Fe–O–Mo/P–O–Fe superexchange pathways.\\
 The nearest-neighbor exchange path ($J_{1}$; 4.74~Å) couples two Fe sites to form a magnetic dimer (Fig.~\ref{XRD}(c)). There are four such dimers per unit cell of \ce{KFMPO}, which are further interconnected via the inter-dimer exchange path ($J_{2}$; 5.03~\AA).  In addition, each Fe sublattice independently forms a trillium lattice, with Fe1–Fe1 and Fe2–Fe2 distances of 6.16~Å ($J_{3}$) and 6.25~Å ($J_{4}$), respectively, corresponding to a chiral network of equilateral triangles (Fig.~\ref{XRD}(d)). The inter-trillium Fe1–Fe2 bond (6.06~Å; $J_{5}$) connects Fe$^{3+}$ ions at inequivalent sites, thus establishing a hypertrillium lattice of three corner-sharing octahedra (Fig.~\ref{XRD}(e)).\\
Overall, the presence of two inequivalent Fe sites results in multiple competing exchange pathways, including dimer ($J_{1}$), inter-dimer ($J_{2}$), trillium ($J_{3}$, $J_{4}$), and hypertrillium ($J_{5}$) interactions in KFMPO (see Table~\ref{EXCH}). Although the conventional Goodenough--Kanamori--Anderson rules are not directly applicable to the extended Fe--O--P/Mo--O--Fe  super-superexchange pathway~\cite{kittel2018introduction}, structural parameters such as bond lengths and bond angles may still provide a qualitative indication of the hierarchy of magnetic interactions.
Accordingly, the exchange interactions in KFMPO are expected to be inequivalent, reflecting the diversity of exchange pathways and local geometries. In particular, the inter-trillium interaction ($J_{5}$)
	is likely dominant, while the trillium ($J_{3}$, $J_{4}$) and inter-dimer ($J_{2}$)
	interactions provide additional competing channels~\cite{PhysRevLett.127.157204,k3lw-2567}.\\ Given the structural complexity of the super-superexchange network, however, the relative strengths of these interactions cannot be uniquely determined from geometry alone. A reliable hierarchy would require microscopic evaluation, for instance through electronic structure calculations, which we leave for future work. We do not consider weaker longer-range interactions in the present discussion. Overall, KFMPO can be qualitatively viewed as a system in which hypertrillium-like connectivity coexists with coupled spin dimers. The competition between these magnetic sub-networks may give rise to thermally driven crossovers in the dominant spin correlations.
  \\\begin{figure*}[t]
	\centering
	\includegraphics[width=\textwidth]{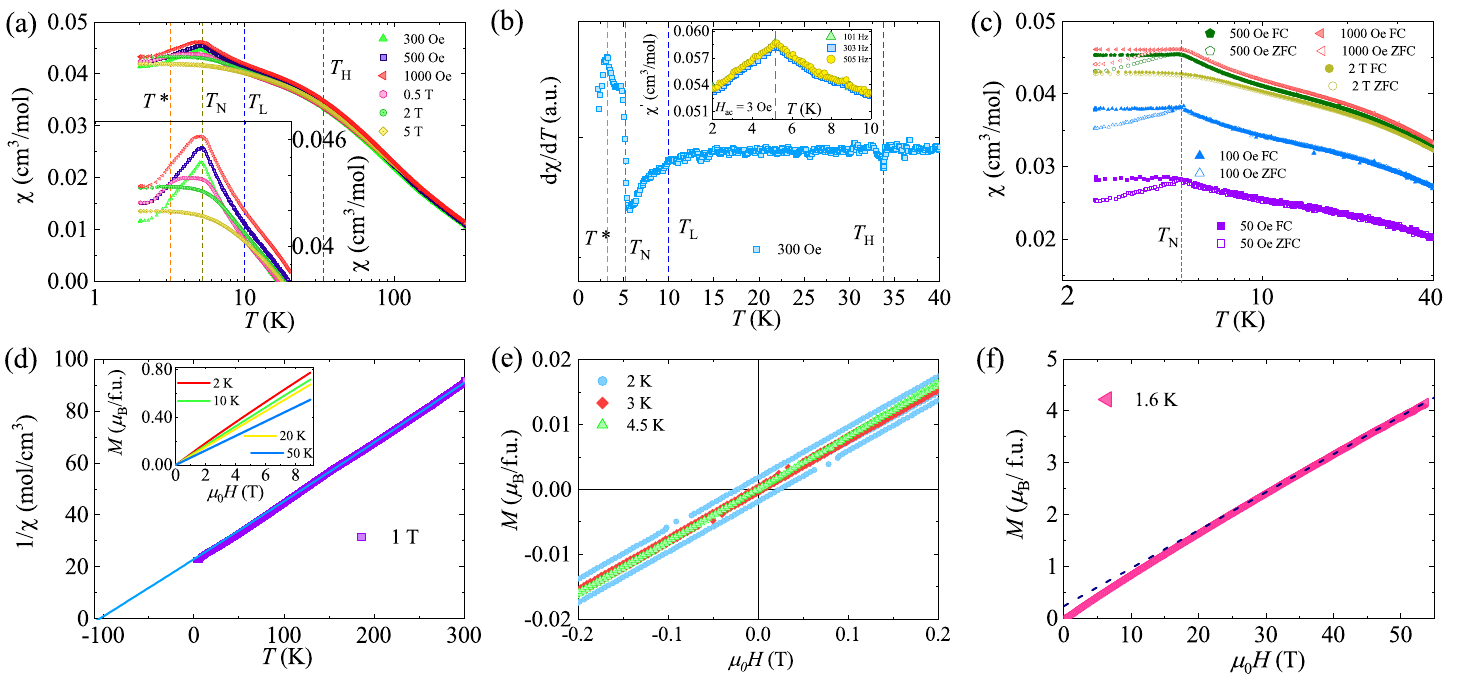}
	\caption{(a) Temperature dependence of the magnetic susceptibility $\chi(T)$ at several magnetic fields. The bottom inset shows an enlarged view of $\chi(T)$ below 20~K. The dashed vertical lines indicate the characteristic temperatures $T_{\mathrm{H}} = 34$~K, $T_{\mathrm{L}} = 10$~K, and $T^{*} = 3.2$~K, which signify changes in spin correlations, while $T_{\mathrm{N}}$ indicates the onset of weak magnetic ordering. (b) Temperature dependence of the derivative of $\chi(T)$ at 300~Oe. The top inset shows the temperature dependence of the real part of the \textit{ac} $\chi(T)$ at three different frequencies. (c) Zero-field-cooled (ZFC) and field-cooled (FC) $\chi(T)$ as a function of temperature at different magnetic fields. (d) Temperature dependence of the inverse $\chi(T)$ together with a Curie--Weiss fit. The top inset depicts the isothermal magnetization as a function of magnetic field for one quadrant at several temperatures. (e)  Hysteresis loops in low-field magnetization measured over five quadrants at three selected temperatures. (f) Isothermal magnetization as a function of magnetic field  up to 55~T. The dashed line is a linear fit at high fields, shown as a guide to the eye. All magnetic susceptibility and magnetization measurements were performed with the applied magnetic field parallel to the [111] direction.}
	\label{Magsus}
\end{figure*}
\subsection{Magnetic susceptibility}
Figure~\ref{Magsus}(a) shows the temperature dependence of the magnetic susceptibility, $\chi(T)$, at several  magnetic fields. Upon lowering the temperature, $\chi(T)$ increases and develops a broad hump around $T_{\rm H} = 34$~K, reflecting the presence of short-range spin correlations driven by the dominant antiferromagnetic exchange interactions~\cite{PhysRevB.95.094427}. On further cooling at $H = 300$~Oe, $\chi(T)$ increases gradually and, unlike in conventional low-dimensional antiferromagnets where it decreases below the broad hump, remains slowly rising down to $T_{\rm L}=10$~K \cite{PhysRevB.95.094427}. This kind of behavior is not compatible with pure antiferromagnetic interactions and likely reflects the influence of the complexity of exchange interactions and competing magnetic correlations present in the present system.\\ \\
Below $T_{\rm L}$, $\chi(T)$ increases rapidly, as evidenced by the temperature derivative  $d\chi/dT$ shown in Fig.~\ref{Magsus}(b), and displays a pronounced $\lambda$-like anomaly at $T_{\rm N} = 5.2$~K for the weak magnetic fields, signaling the onset of  magnetic order (see  the inset of Fig.~\ref{Magsus}(a))~\cite{PhysRevB.95.094427}. 
 On further cooling below $T_{\rm N}$, $\chi(T)$ decreases sharply and eventually approaches a saturation-like behavior below 2.5~K. In addition, the temperature derivative $d\chi/dT$ at $H$ = 300 Oe reveals a secondary anomaly at $T^{*} = 3.2$~K (Fig.~\ref{Magsus}(b)), which we attribute  tentatively to a spin reorientation arising from residual magnetic interactions at low temperatures.
 \\ 
To further elucidate the nature of the anomaly at $T_{\mathrm{N}}$,  $\chi(T)$ measurements were carried out under several applied magnetic fields. The broad feature at $T_{\rm H}$ remains essentially unchanged up to the highest measured field. In contrast, below $T_{\rm L}$, $\chi(T)$ is initially enhanced and then  is suppressed, while the $\lambda$-like anomaly becomes progressively broadened with increasing field (see the inset of Fig.~\ref{Magsus}(a)), indicating the weak canted  antiferromagnetic character of the magnetic ordering, which is strengthened by the competition between exchange interactions and the Zeeman energy under weak applied magnetic fields. For $\mu_{0}H \geq 2$~T, the anomaly at $T_{\rm N}$ is strongly suppressed, and the corresponding $\chi(T)$ decreases with further increasing field at low temperatures, suggesting that the applied magnetic field  gradually polarizes the weak ferromagnetic moments~\cite{PhysRevB.95.094427}.\\
All features observed in $\chi(T)$ are further confirmed by its temperature derivative, $d\chi/dT$, as presented in Fig.~\ref{Magsus}(b). The dashed vertical lines indicate the same characteristic temperatures identified in Fig.~\ref{Magsus}(a), thereby confirming the consistency of the observed anomalies. To obtain additional evidence for canted antiferromagnetic ordering, the $ac$ $\chi(T)$ measurements were performed. As shown in the inset of Fig.~\ref{Magsus}(b), a clear anomaly is observed at $T_{\mathrm{N}} = 5.2$~K. Notably, the absence of any frequency dependence of this anomaly suggests that the transition at $T_{\mathrm{N}}$ is not associated with conventional spin-glass-like  behavior, but reflects the presence of canted antiferromagnetic   ordering~\cite{Goi2000}. Although the present $ac$ $\chi(T)$ data are limited to relatively low frequencies, our zero-field $\mu$SR results (sec.~\ref{MUSR}) further rule out a conventional spin freezing at $T_{\mathrm{N}}$, supporting the presence of  weak magnetic order.\\ \\\begin{figure*}
	\centering
	\includegraphics[ width=\textwidth]{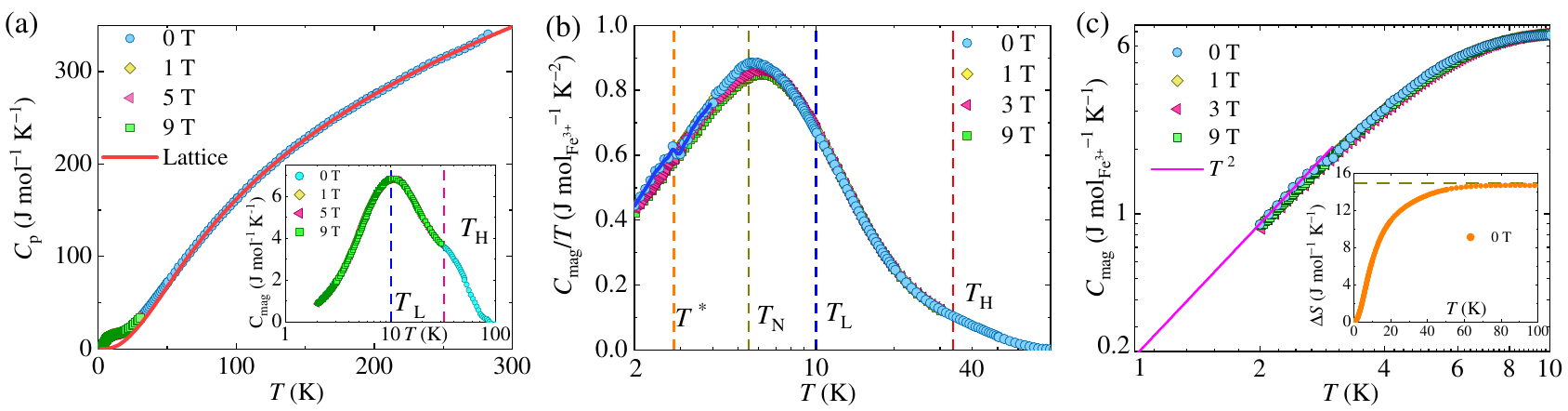}
	\caption{(a)Temperature dependence of the specific heat  at several magnetic fields. The solid orange line represents the lattice contribution, as described in the text. The inset shows the magnetic specific heat ($C_{\rm mag}$) as a function of temperature for different fields. An inflection point around $T_{\rm H} \approx 34 $ K and a broad maximum at $T_{\rm L}\approx 10$ K are indicated by the red and blue dashed vertical lines, respectively.
		(b) Temperature dependence of $C_{\rm mag}/T$ that highlights two additional features; a broad peak of weak magnetic ordering at $T_{\rm N}\approx 5.2$ K and a crossover to distinct spin correlations below $T^{*} = 3.2$ K, indicated by dark yellow and orange vertical lines, respectively. The solid blue line below 4 K guides the eye.
		(c) Temperature- and field-dependent $C_{\rm mag}$, which follows a quadratic temperature dependence below $T^{*}$. The inset depicts the entropy change as a function of temperature where the dashed horizontal line marks the expected entropy (\textit{R} ln6) for a $S=5/2$ system.  } {\label{Specific}}
\end{figure*}  Furthermore, zero-field-cooled (ZFC) and field-cooled (FC) dc $\chi(T)$ measurements were conducted at magnetic fields ranging from 50~Oe to 2~T (see Fig.~\ref{Magsus}(c)). At a low field of 50~Oe, a clear bifurcation between the ZFC and FC $\chi(T)$ curves is observed below $T_{\mathrm{N}}$, indicative of domain-wall dynamics enhanced by magnetic anisotropy and weak disorder. The absence of a Curie-like upturn in low-temperature $\chi(T)$, unlike in other Fe-based compounds~\cite{PhysRevB.81.024506}, indicates negligible contributions from paramagnetic impurity defects in KFMPO. With increasing magnetic field, the magnitude of $\chi(T)$ increases, while the ZFC–FC splitting remains robust and persists up to fields of approximately 1000~Oe. Upon further increasing the field to 2~T, $\chi(T)$ decreases and the ZFC–FC bifurcation disappears, indicating the suppression of domain-related effects due to the polarization of weak canted moments by the applied field. This field evolution clearly reflects the presence of weak canted antiferromagnetic ordering~\cite{PhysRevB.109.184432} or DM-induced complex magnetic structure with weak ferromagnetic component~\cite{PhysRevB.95.094427,Goi2000} that is initially strengthened at low to intermediate fields but ultimately suppressed at sufficiently high fields due to spin polarization. In contrast, for typical spin-glass or antiferromagnetic systems, an applied magnetic field generally suppresses the $\chi(T)$ anomaly~\cite{doi:10.1143/JPSJ.63.3122,PhysRevB.110.184402}. \\
Shown in Fig.~\ref{Magsus}(d) is the temperature dependence of the inverse magnetic susceptibility, $1/\chi(T)$, which follows a Curie–Weiss (CW) law, $\chi(T) = \chi_{0} + C/(T - \theta_{\rm CW})$, in the temperature range 150–300~K (solid line in Fig.~\ref{Magsus}(d)). The fitting yields a temperature-independent magnetic susceptibility $\chi_{0} = -3.77 \times 10^{-4}$~cm$^{3}$/mol, a Curie constant $C = 4.60$~cm$^{3}$ K/mol, and a CW temperature $\theta_{\rm CW} = -104$~K. The effective magnetic moment estimated from the Curie constant, $\mu_{\rm eff} = \sqrt{8C} = 6.06~\mu_{\rm B}$, is close to the expected theoretical value $\mu_{\rm eff}^{\rm the} = 5.9~\mu_{\rm B}$ for free Fe$^{3+}$ ions with spin $S = 5/2$. The large negative CW temperature suggests the presence of strong dominant antiferromagnetic interactions, as inferred from the Goodenough–Kanamori–Anderson rules~(see Table~\ref{EXCH}).\\
Below 150~K, the experimental data progressively deviate from the high-temperature CW behavior, reflecting the development of magnetic correlations.  Notably, the estimated frustration parameter, $f = |\theta_{\mathrm{CW}}|/T_{\mathrm{N}} \approx 20$, implies that the system is  highly frustrated.\\
The field dependence of the isothermal magnetization is shown in the inset of Fig.~\ref{Magsus}(d). It manifests an almost linear increase with applied fields up to 9~T, without any indication of  saturation of local moments, which further confirms the dominance of strong antiferromagnetic interactions in this system. However, as shown in Fig.~\ref{Magsus}(e), at temperatures $T \leq 4.5$~K, the isothermal magnetization curves exhibit a weak hysteresis, providing additional evidence for the presence of  weak ferromagnetic interactions or canted magnetic moments coexisting with strong antiferromagnetic interactions~\cite{PhysRevB.95.094427}.  It is worth noting that, similar to the present compound, signatures of weak canted antiferromagnetic order coexisting with strong antiferromagnetic interactions have also been observed in the double trillium compound K$_2$FeSn(PO$_4$)$_3$~\cite{lmsf-73hn}. \\
To further investigate whether the  magnetization exhibits a \( \tfrac{1}{3} \) plateau at a field \(\mu_{0}H \simeq 0.8J\) (where \(J\) denotes the nearest-neighbour exchange interaction in  Kelvin), similar to the behaviour observed in the \(S=5/2\) single trillium‑lattice compound Na[Mn(HCOO)$_{3}$]~\cite{PhysRevLett.128.177201}, we extended the isothermal magnetization measurements to high magnetic fields up to 55\,T, as shown in Fig.~\ref{Magsus}(f).
 The absence of a one-third magnetization plateau indicates that additional magnetic interactions arising from the double trillium lattice geometry play an important role in this system~\cite{PhysRevLett.128.177201}. At 55~T, the magnetization reaches approximately $4~\mu_{\rm B}$ per formula unit.  This corresponds to approximately 40 \% of the full saturation magnetization, implying that the fully polarized state would be reached at fields on the order of $\sim$140 T ($\sim$ 55 T/0.40). This saturation field is comparable to that inferred from the magnitude of the CW temperature $|\theta_{\mathrm{CW}}|$.
\subsection{Specific heat}
To gain further insight into the features observed in $\chi(T)$, specific heat measurements were performed at several magnetic fields down to 2 K. Figure~\ref{Specific}(a) depicts the temperature dependence of the specific heat ($C_{\rm p}(T)$)  at various magnetic fields. Despite the weak magnetic ordering around $T_{\rm N} = 5.2$ K inferred from $\chi(T)$,  $C_{\rm p}(T)$ shows no $\lambda$-type anomaly at this temperature. This absence indicates that strong antiferromagnetic spin correlations dominate the thermodynamic response and effectively mask the specific heat signature of the weak  ordering~\cite{PhysRevB.95.094427}.\\\\ To reveal such a weak feature, it is worth extracting the magnetic contribution in $C_{\rm p}(T)$. Accordingly, the $C_{\rm p}(T)$ data were fitted in the temperature range 100–220 K using a lattice model (orange line in Fig.~\ref{Specific}(a)) comprising one Debye and three Einstein terms, i.e., \begin{equation}
\begin{split}\label{DE}
C_{\rm p}^{\rm ph}(T) = C_{\rm D} \left[ 9R \left(\frac{T}{\theta_{\rm D}}\right)^3 
\int_0^{\theta_{\rm D}/T} \frac{x^4 e^x}{(e^x-1)^2} \, dx \right] \\
+ \sum_{i=1}^{3} C_{\rm E_{i}} \left[ R \left(\frac{\theta_{\rm E_{i}}}{T}\right)^2 
\frac{\exp\left(\frac{\theta_{\rm E_{i}}}{T}\right)}{\left[\exp\left(\frac{\theta_{\rm E_{i}}}{T}\right)-1\right]^2} \right],
\end{split}
\end{equation}
where $\theta_{\rm D}$ represents the Debye temperature, $\theta_{\rm E_is}$ correspond to the characteristic Einstein temperatures of the three optical modes, and $R$ is the molar gas constant. The best fit was obtained by fixing the weighting coefficients to $C_{\rm D}=3$, $C_{\rm E_1}=15$, $C_{\rm E_2}=18$, and $C_{\rm E_3}=21$, yielding a Debye temperature of $\theta_{\rm D}=162(10)$ K and Einstein temperatures of $\theta_{\rm E_1}=583(4)$ K, $\theta_{\rm E_2}=264(7)$ K, and $\theta_{\rm E_3}=1306(5)$ K. Notably, the weighting coefficients closely follow the expected distribution of three acoustic modes and $(3n-3)$ optical phonon branches, where $n$ is the total number of atoms in \ce{KFMPO} \cite{PhysRevB.110.184402}.
\\
The magnetic specific heat, $C_{\rm mag}(T)$, obtained after subtraction of the lattice contribution, is shown in the inset of Fig.~\ref{Specific}(a). It starts to increase below 100 K, reflecting the growth of magnetic correlations consistent with the estimated high-temperature CW temperature ($|\theta_{\rm CW}|\approx104$ K). At temperatures slightly above $T_{\rm H}=34$ K, where  $\chi(T)$ exhibits a broad feature, $C_{\rm mag}(T)$ shows a weak hump. Below this temperature, $C_{\rm mag}(T)$ increases again and reaches a broad maximum near $T_{\rm L}=10$ K. The weak hump at $T_{\mathrm{H}}$, followed by a subsequent maximum at $T_{\mathrm{L}}$, is associated with two distinct onset temperature scales of short-range magnetic order, each pertaining to distinct sublattices associated with the anisotropic hypertrillium network and the coupled spin dimers. We further note that $T_{\mathrm{H}} \sim 0.32\,|\theta_{\mathrm{CW}}|$, consistent with the typical relation between the characteristic temperature scale of short-range magnetic correlations and the CW temperature in frustrated antiferromagnets~\cite{PhysRevLett.134.226701}. The evolution of short-range spin correlations well above $T_{\rm H}$ is further elucidated by ESR experiments, as discussed in  Sec.~\ref{ESR}. \\ With further lowering the temperature below $T_{\rm L}$ and through $T_{\rm N}$, $C_{\rm mag}(T)$ decreases smoothly without showing a sharp anomaly, although a noticeable change in slope is observed. To examine this behavior more closely, we plot $C_{\rm mag}(T)/T$ as a function of temperature in Fig.~\ref{Specific}(b). Interestingly, a broad maximum appears around $T_{\rm N}$, which is attributed to weak magnetic ordering, consistent with similar observations in other compounds~\cite{PhysRevB.95.094427}. Furthermore, the weak field dependence of this broad maximum is consistent with the field-induced evolution of the $\chi(T)$ anomaly at $T_{\mathrm{N}}$ (see Fig.~\ref{Magsus}(a)). In addition, a weak anomaly is observed near $T^{*}$, reflecting a slight spin reorganization driven by residual magnetic interactions. Below $T^{*}$, $C_{\rm mag}(T)$ exhibits a quadratic-like temperature dependence, possibly associated with local spin-singlet excitations in the present  $S=5/2$ double trillium lattice compound or exotic gapless excitations~\cite{PhysRevLett.84.2957}. Although our data are limited to temperatures down to 2 K, it has been observed that other $S=5/2$ double trillium compounds exhibit a quadratic temperature dependence of $C_{\rm mag}(T)$ at low temperatures~\cite{lmsf-73hn,10.1063/5.0096942}.\\ \\By extrapolating the quadratic low-temperature behavior of $C_{\rm mag}$ below 2 K, the magnetic entropy ($\Delta$\textit{S}) was calculated by integrating $C_{\rm mag}(T)/T$, and plotted in the inset of Fig.~\ref{Specific}(c). The entropy saturates at temperatures comparable to the CW temperature, and its magnitude is in good agreement with the expected value $R\ln(2S+1)$ for an $S=5/2$ system with Fe$^{3+}$ ions. At $T_{\rm N}$, only about 13 \%
 of the total magnetic entropy is released, with the remaining entropy recovered at higher temperatures owing to short-range spin correlations. Although strong antiferromagnetic exchange interactions dominate, the origin of the weak canted antiferromagnetic-like magnetic ordering remains unclear~\cite{B002076L}. It may originate from sublattice ordering, interactions among inequivalent magnetic sites without a well-defined sublattice structure, or from the interplay of strong Heisenberg exchange and weak DM interactions intrinsic to trillium lattices~\cite{B002076L,PhysRevB.95.094427,Khatua_2025}. Future neutron diffraction experiments will be essential to unambiguously determine the magnetic ordering structure. 
\begin{figure}[b]
	\centering
	\includegraphics[width=0.5\textwidth]{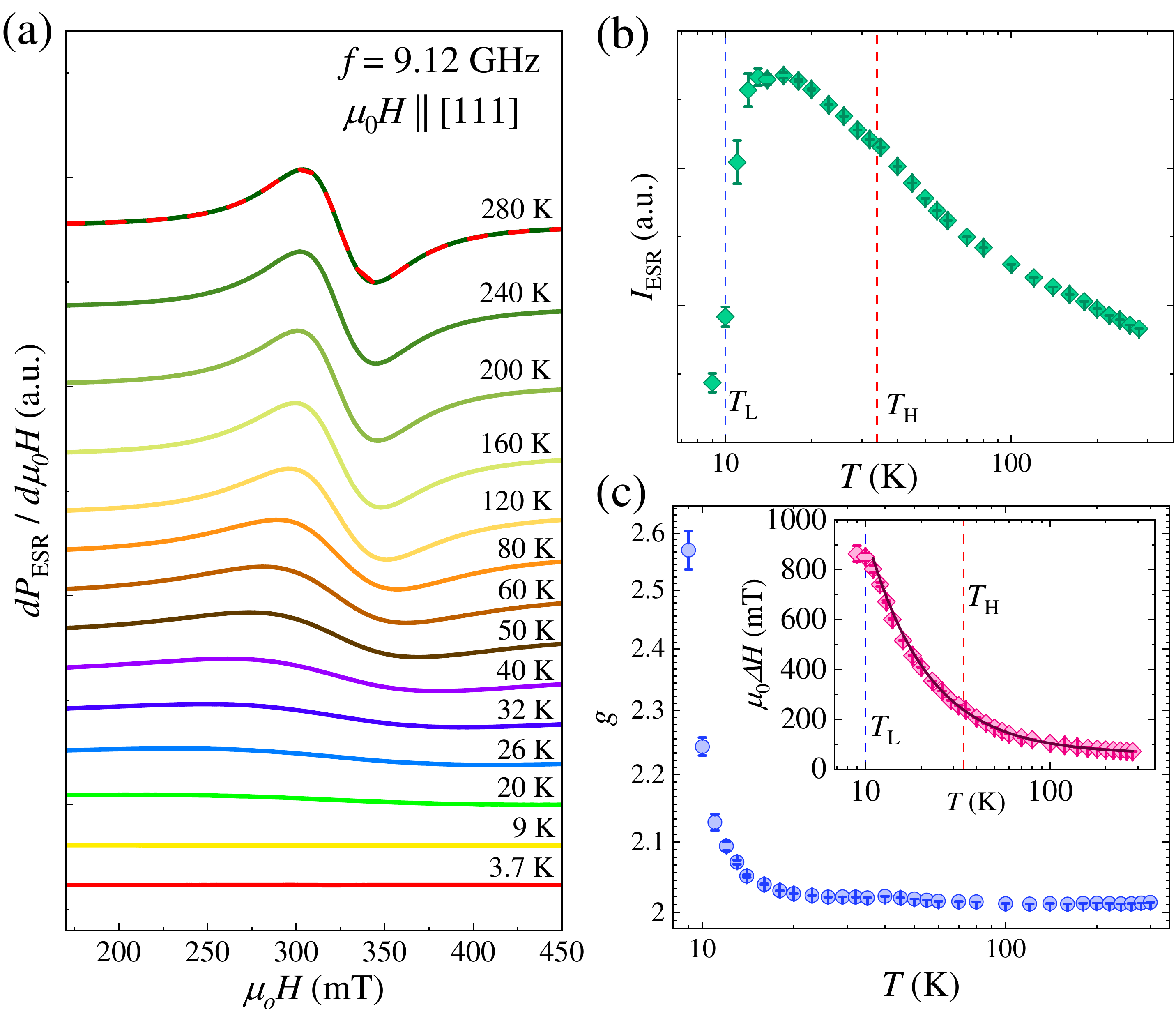}
	\caption{(a) Selected ESR spectra at $f$ = 9.12 GHz at various temperatures with the magnetic field applied to the [111] crystal orientation. Spectra are vertically offset for clarity. The dashed red line on the spectrum at $T$ = 280 K represents the fitting using a derivative Lorentzian function. Temperature dependence of (b) the integrated ESR intensity, (c) the $g$-factor, and the peak-to-peak linewidth (in the inset) obtained from the fittings of the ESR spectra shown in (a). The black solid curve in the inset corresponds to a power-law fit (see the text in detail). (b) and (c) are semi-log plots. The dashed vertical lines in (b) and (c) are at $T_{\rm H}$ and $T_{\rm L}$, as in Fig.~\ref{Specific}.
	}
	\label{ESRfig}
\end{figure}
\subsection{Electron spin resonance}\label{ESR}
To elucidate a thermal evolution of spin correlations, X-band   
ESR spectra of \ce{KFMPO} were measured at $f$ = 9.12 GHz  with magnetic field applied along the [111] crystal orientation (Fig.~\ref{ESRfig}(a)). A single resonance feature is observed at $T$ = 280 K and it gets broader as the temperature decreases to $T_{\rm L}$. Below $T = 9$ K, close to $T_{\rm L}$ where $C_{\rm mag}(T)$ exhibits a broad maximum, the signal becomes excessively broadened over the entire field range, and no well-defined resonance is observed, indicating the development of critical spin fluctuations. The observed resonance feature is well fitted using a derivative Lorentzian line profile. \\
Temperature-dependent ESR parameters, including the integrated intensity, $g$-factor, and peak-to-peak linewidth were extracted from fits to  the ESR spectra shown in Fig.~\ref{ESRfig}(a) and are presented in Fig.~\ref{ESRfig}(b) and Fig.~\ref{ESRfig}(c) as semi-log plots. The integrated intensity shows an increasing behavior down to $T$ = 18 K (Fig.~\ref{ESRfig}(b)). Following a broad maximum near 
15 K, the integrated intensity decreases abruptly below 12 K ($\sim$$T_{\rm L}$), while $\chi(T)$ continues to rise down to 2 K. Specifically, the marked deviation of the integrated resonance intensity from $\chi(T)$ implies the partial wipeout of the ESR signal as the temperature is lowered below $T_{\rm H}$. The wipe-out effect arises from the resonance linewidth exceeding the instrumental bandwidth of the ~9 GHz spectrometer. Further, the loss of intensity indicates that a majority of the spin ensemble is participating in fluctuations too fast or too broad to be captured within the accessible magnetic field range~\cite{Goi2000}. Despite the substantial wipe-out, the surviving X-band resonance offers a frequency-selective probe, capturing the slower fluctuations within the broad temporal distribution of the spin ensemble.\\ The buildup of the internal field is also unveiled in the shift of $g$-value (Fig.~\ref{ESRfig}(c)). At high temperatures above $T$ = 60 K, the $g$-value is constant, $g$ = 2.01, slightly larger than the typical value 
$g$ = 2.00 expected for half-filled Fe$^{3+}$ ($S$ = 5/2) ions~\cite{kittel2018introduction}. Upon cooling from 60~K to 20~K, the $g$ value increases gradually to about 2.02.
At lower temperatures, especially below 20~K, the $g$ value increases abruptly and reaches $g = 2.56$ at $T = 9$~K. Although the low-$T$ $g$-factor should be taken as an effective parameter, it can serve as qualitative indicators of the progressive slowing down of spin fluctuations. In this sense, the sharply increasing $g$-factor indicates the development of strong, critical-like spin fluctuations and may signal the incipient magnetic ordering~\cite{lmsf-73hn,Wulferding_2012}.
 \\ \\
The temperature-dependent evolution of spin correlations is more clearly revealed by the temperature-dependent ESR linewidth broadening (see the inset of Fig.~\ref{ESRfig}(c)). The resonance linewidth increases monotonically upon cooling over the entire measured temperature range, except below 9 K, where it becomes indeterminable due to a significant wipeout effect. Such a pronounced temperature dependence of the ESR linewidth persisting up to $T = 280$~K, corresponding to approximately $3|\theta_{\mathrm{CW}}|$, is a characteristic feature of frustrated antiferromagnets, reflecting the persistence of short-range spin correlations far above the CW temperature~\cite{Wulferding_2012}. For further quantification of the linewidth broadening, the linewidth data are fitted using a power-law function, $\mu_{0}$$\Delta H \propto T^{-p}$. The power-law fit yields an exponent $p$ = 1.27(4) all the way down to $T_{\rm L}$, indicating that the $T_{\rm L}$ anomaly is captured by the X-band frequency window. The pronounced linewidth broadening and the relatively large exponent reflect rapid spin relaxation driven by strong spin fluctuations~\cite{lmsf-73hn}. It is noteworthy that the exponent value is slightly smaller than the reported value of $p$ = 1.5 for the 3D magnet case~\cite{h2220570227}, suggesting the presence of residual short-range correlations.
\begin{figure*}
	\centering
	\includegraphics[width=0.77\textwidth]{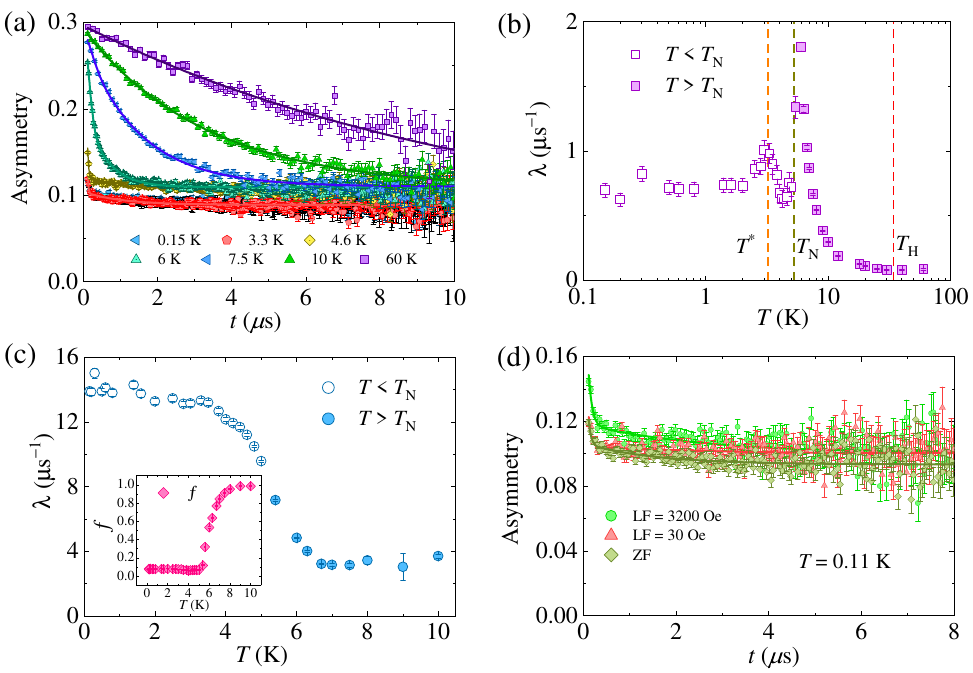}
	\caption{(a) Time evolution of the zero-field muon spin asymmetry at selected temperatures.
		(b) Temperature dependence of the slow Lorentzian muon spin relaxation rate, $\lambda$, plotted in a semi-logarithmic scale. The vertical dashed line corresponds to  the same characteristic temperature as indicated in Fig.~\ref{Magsus}(a).
			(c) Temperature dependence of the fast Gaussian relaxation rate, $\sigma$. The inset illustrates the temperature dependence of the Lorentzian fraction.
		(d) Time evolution of the muon spin asymmetry at longitudinal magnetic fields for $T = 110$~mK.
	}
	\label{Musrfig}
\end{figure*} 
\subsection{Muon spin relaxation}\label{MUSR}
To further validate the presence of weak magnetic order and to probe the associated spin dynamics,  $\mu$SR measurements were performed on randomly aligned single crystals of \ce{KFMPO}. Figure~\ref{Musrfig}(a) shows the time evolution of the zero-field (ZF) muon spin asymmetry down to 150~mK.\\ In polycrystalline samples, the coexistence of static and dynamic local electronic spin moments typically gives rise to a well-defined asymmetry pattern, characterized by coherent oscillations with an initial amplitude of $\approx$ $2/3$ of the total asymmetry at short times and a non-oscillatory $1/3$ tail at late times originating from spin components parallel to the internal magnetic field below the transition temperature. In the present system, however, no coherent oscillations are observed down to  $T = 150 \ \text{mK} \ll T_{\rm N}$. Instead, the muon spin asymmetry exhibits  a fast early-time relaxation at low temperatures, followed by a slowly relaxing component that persists to late times, indicating the presence of quasistatic local magnetic fields and slow spin dynamics in \ce{KFMPO}. Furthermore, if the slow spin dynamics originates from a conventional spin-glass-like state, one would expect a recovery of the $1/3$ tail in the muon asymmetry, which is absent in the titled compound~\cite{PhysRevB.31.546}.\\
To gain insight into the temperature evolution of the spin dynamics, the $\mu$SR spectra were fitted using the function
\begin{equation}
A(t) = A_{0} e^{-\lambda t} + A_{\rm bg},
\end{equation}
where $A_{0}$ is the initial asymmetry,  $A_{\rm bg}$ represents the background contribution, and $\lambda$ is the muon spin relaxation rate, which characterizes the cooperative dynamics of fluctuating local moments.
To minimize the number of fitting parameters, both $A_{0}$ and $A_{\rm bg}$ were fixed at $A_{0}=0.293$ and $A_{\rm bg}=0.079$, respectively. This exponential form provides a satisfactory description of the data for $T > T_{\mathrm{L}}$. For $T \leq T_{\mathrm{L}}$, however, the asymmetry exhibits a clear non-exponential behavior, indicating the presence of an additional relaxation channel. After examining several alternative models, including single- and stretched-exponential forms, we find that the spectra are best described by a two-component relaxation function, in which the early-time relaxation is captured by a Gaussian term and the late-time behavior by a Lorentzian component. Accordingly, the $\mu$SR spectra  in the $T$-range from 10 K to 150~mK were fitted using~\cite{l7gq-cc96,PhysRevB.70.020401}
 \begin{equation}\label{REV}
 A(t) = A_{0}\left[ f e^{-\lambda t} + (1-f)e^{-(\sigma t)^2} \right] + A_{\rm bg},
 \end{equation}
 where $f$ measures the fractional amplitude of the slow Lorentzian relaxation component.\\
 The temperature dependence of $\lambda(T)$ is shown in Fig.~\ref{Musrfig}(b), which is nearly temperature independent above 30 K ($\sim$ $T_{\mathrm{H}}$), indicating a  motional narrowing regime with exchange fluctuations of electronic spins. Upon cooling below 30 K, $\lambda(T)$ increases gradually and exhibits a rapid increase  as the system approaches $T_{\mathrm{N}}$. The resulting critical divergence of $\lambda(T)$ toward $T_{\mathrm{N}}$ provides microscopic evidence for the onset of weak magnetic ordering. Note that in ordered magnets with quasistatic local moments, the slow late-time relaxation captures critical-like fluctuations associated with magnetic ordering, while the fast transverse component is largely insensitive to dynamical effects~\cite{PhysRevB.28.6209}. \\Below $T_{\mathrm{L}}$, an increase in $\lambda(T)$ is observed, which may reflect the slowing down of spin fluctuations within the $\mu$SR time window upon approaching the magnetic transition. However, given the limitations of pulsed $\mu$SR measurements, particularly the loss of early-time asymmetry, any quantitative analysis of critical behavior, such as extracting a reliable critical exponent, is not well constrained in the present case~\cite{PhysRevB.83.104404,PhysRevB.61.4060}. Notably, the GHz spin dynamics in this temperature range exhibits signs of saturation [see inset, Fig.~\ref{ESRfig}(c)].  \\ It is worth noting that the use of open symbols for data below $T_{\rm N}$ (see Fig.~\ref{Musrfig}(b) and Fig.~\ref{Musrfig}(c)) reflects the higher modeling uncertainty imposed by the limited time resolution of the pulsed muon source, although Eq.~\ref{REV} remains the most suitable description.
Following the sharp divergence of $\lambda(T)$, it initially decreases and then rises again upon further cooling, forming a weak shoulder peak at $T^{*} = 3.2$~K. Consistent with thermodynamic measurements, this feature is attributed to  a reorientation of the ordered magnetic moments. Furthermore, a weak crossover-like behavior in the late-time asymmetry between 3.3 K and 0.15 K is observed in the zoomed view of the tail region (not shown here),  further indicating a change in spin dynamics at low temperatures.    Notably, the temperature-independent behavior of $\lambda(T)$ below 2~K indicates persistent spin dynamics even in the magnetically ordered state.\\ 
The fast Gaussian relaxation rate $\sigma(T)$ at early times, arising from quasistatic local moments, is plotted in Fig.~\ref{Magsus}(c). It remains nearly temperature independent down to 6.5~K, below which it starts to increase, indicating a slowing down of spin fluctuations above $T_{\rm N}$. Below 4~K, $\sigma(T)$ becomes only weakly temperature dependent and exhibits a plateau-like behavior, suggesting the development of predominantly quasistatic internal fields. We note that at $\sigma(T) \sim 14~\mu\mathrm{s}^{-1}$, the relaxation approaches the time-resolution limit of the pulsed muon source at ISIS, and the extracted $\sigma(T)$ mainly reflects the presence of a very fast relaxing component beyond experimental resolution at low temperatures. Similar to the present system, a closely related compound \ce{K2FeSn(PO4)3}~\cite{lmsf-73hn}, measured at PSI using a continuous muon beam, shows a pronounced early-time Gaussian relaxation arising from quasistatic local moments.
\\
The inset of Fig.~\ref{Musrfig}(c) shows the temperature dependence of the Lorentzian fraction $f$, which represents the relative weight of the dynamic relaxation component in the $\mu$SR asymmetry. At high temperatures, 
$f$ is close to one, indicating that the muon spin relaxation is dominated by dynamically fluctuating local fields. Below $\sim$8 K, $f$ decreases rapidly, signaling the gradual development of quasistatic internal fields. At the lowest temperatures, $f$ approaches a small but finite value, suggesting that a fraction of spins remains dynamic even in the magnetically ordered state, consistent with persistent spin dynamics~\cite{lmsf-73hn}.\\
To further elucidate the origin of the Gaussian relaxation observed in the present compound,  it is worthwhile to estimate the associated width of the internal field distribution experienced by the muons. Using the mean-field expression for the average exchange interaction, we obtain
$J = 3k_{\rm B}\theta_{\rm CW}/2zS(S+1) = 2.97~\text{K}$
for $z = 6$ and $S = 5/2$. This yields an exchange fluctuation rate
$\nu = \frac{\sqrt{z}\, J S}{\hbar} \approx 2.4 \times 10^{12}~\text{s}^{-1}
$
\cite{PhysRevLett.73.3306}. Combining this with the high-temperature relaxation rate in the motional narrowing regime
$\lambda = 0.087~\mu\text{s}^{-1}$, we estimate the internal field
distribution width
$
\Delta/\gamma_\mu
= \sqrt{\nu \lambda/2}/{\gamma_\mu}
\approx 3800~\text{Oe}
$ for \ce{KFMPO}. This implies that a longitudinal magnetic field exceeding $\sim$3800 Oe is required to fully decouple the muon spins from the quasistatic internal fields responsible for the early-time Gaussian relaxation. In agreement with this scenario, Fig.~\ref{Musrfig}(d) demonstrates that only weak decoupling is observed up to an applied field of 3200~Oe ($\ll \Delta/\gamma_\mu$) at 110~mK. The persistence of substantial relaxation even at 3200~Oe  indicates that, in addition to quasistatic internal fields, a dynamical relaxation channel remains active at low temperatures. The precise origin of the internal field distribution remains unclear; however, they likely arise from dipolar and transferred hyperfine couplings to the antiferromagnetic sublattice moments, not only from the small net moment associated with the weak magnetic order~\cite{Goi2000,k3lw-2567}. Future $\mu$SR investigations employing a continuous muon beam, combined with field-dependent measurements, would provide important insights into the  magnetic ground state and its evolution under applied magnetic fields.
\begin{table*}[t]
	\caption{Comparison of representative Fe$^{3+}$-based trillium compounds. The table summarizes the possible sources of disorder, Curie--Weiss temperature $\theta_{\rm CW}$, magnetic ordering temperature $T_{\rm N}$, and the frustration parameter $f = |\theta_{\rm CW}|/T_{\rm N}$.}
	\label{tab:comparison}
	\begin{ruledtabular}
		\begin{tabular}{lcccccc}
			Compound ($P2_{1}3; S= 5/2$) & Source of Disorder & $\theta_{\rm CW}$ (K) & $T_{\rm N}$ (K) & $C_{\rm mag}^{\rm broad \ \  peak}$ (K) & $f $ \\
			\hline
			KSrFe$_2$(PO$_4$)$_3$ ($a = 9.80$ \AA) 
			& K/Sr site mixing 
			& $\sim -76$ 
			& $\sim 3.4 \ \ ?$ & $\sim$9  
			& $\sim 22$
			&  \cite{k3lw-2567,Khatua_2025} \\
			
			K$_{2}$FeSn(PO$_{4}$)$_{3}$ ($a$ = 9.91 \AA )
			& Fe/Sn site mixing 
			& $\sim -43$ 
			& $\sim 2$ & $\sim$6.3
			& $\sim 21$
			& \cite{lmsf-73hn}\\
			Pb$_{1.5}$Fe$_{2}$(PO$_{4}$)$_{3}$ ($a$ = 9.91 \AA )
			&  Pb-site deficiency
			& $\sim -68$ 
			& $-- $  &$\sim$ 5
			& $--$
			& \cite{BOYA2025173302}\\
			K$_2$Fe$_2$(MoO$_4$)(PO$_4$)$_2$ ($a$ = 10.05 \AA )
			& Mo/P site mixing 
			& $\sim -104$ 
			& $\sim 5.2$ & $\sim$10
			& $\sim 20$
			& \textbf{This work}  \\
		\end{tabular}
	\end{ruledtabular}
\end{table*}
\section{Discussion} In real materials, a certain degree of disorder is unavoidable owing to intrinsic imperfections such as vacancies, site mixing, and lattice distortions, and can therefore play an important role in shaping the magnetic ground state~\cite{annurevth}. As summarized in Table~\ref{tab:comparison}, various types of structural disorder are commonly encountered in Fe$^{3+}$-based trillium compounds. Despite the differing nature of the disorder, the magnetic ordering temperatures remain confined to a relatively narrow range ($T_{\rm N} \sim 2$--5~K), showing only minor variations. This suggests that, in highly frustrated, high-spin ($S$ = 5/2) magnets with a hierarchy of exchange interactions, the relationship between disorder and magnetic ordering is multifaceted. While disorder can perturb magnetic ordering, it may also play a cooperative role in selecting or stabilizing the ordered state. This complexity is consistent with recent proposals that disorder can even induce unconventional magnetic states, such as chiral order in kagome systems~\cite{yyzf-jjc6}.  \\
Interestingly, neutron diffraction measurements on KSrFe$_2$(PO$_4$)$_3$~\cite{10.1063/5.0096942} did not reveal clear long-range magnetic order, whereas a later bulk $ac$ magnetic susceptibility study revealed a weak magnetic anomaly near $T_{\rm N}$~\cite{Khatua_2025}, similar to those observed in K$_{2}$FeSn(PO$_{4}$)$_{3}$~\cite{lmsf-73hn} and the present system KFMPO. This apparent discrepancy points to an unconventional magnetic ground state in which static long-range order is either strongly suppressed or spatially inhomogeneous.  Another possibility is that the weak magnetic anomaly does not originate from a conventional antiferromagnetic ordered state, but rather from ferromagnetically aligned moments with a substantially reduced ordered moment due to magnetic frustration, as observed in ZnYb$_2$S$_4$~\cite{JPSJ.95.074706}.  In this context, further microscopic investigations incorporating the possible role of DM interactions would be highly valuable for elucidating the origin of such weak ordering tendencies in these materials~\cite{PhysRevB.95.094427,Lane_2025}.    \\ 
Moreover, a common feature emerging across all known $S=5/2$ trillium systems is that the observed magnetic ordering is unlikely to arise from simple sublattice ordering alone. Instead, it may originate from a small imbalance or canting of magnetic moments between two interpenetrating sublattices, resulting in a weak ferrimagnetic or canted antiferromagnetic state~\cite{BATTLE198616,PhysRevB.109.184432}. Such behavior is reminiscent of recent discussions in frustrated magnets, where a weak net magnetization can emerge from competing interactions and geometrical frustration~\cite{Gonzalez2025}. Notably, ferromagnetic-like ordering is often observed in systems containing two inequivalent magnetic sublattices, as reported for \ce{$\alpha$-Li3Fe2(PO4)3}~\cite{Goi2000} and \ce{Li3Fe2(AsO4)_{3-x}(PO4)_{x}}~\cite{B002076L}.\\
In several previously reported Fe-based trillium candidates, it remained unclear whether the low-temperature anomaly originated from weak magnetic ordering or from an unconventional freezing process~\cite{k3lw-2567}. In contrast, the present system provides clearer evidence for a genuine magnetic transition, as the magnetic susceptibility increases with applied field while the bifurcation between the ZFC and FC curves remains intact. Such behavior was not clearly established in the earlier compounds and suggests a more complex weak magnetically ordered state in such high-spin trillium based compounds. Moreover,  the magnetically ordered state in KFMPO is corroborated by specific heat and $\mu$SR measurements. Additionally, below $T_{\rm N}$, a secondary anomaly is observed, which is likely associated with a spin reorientation and is consistent with the thermodynamic data. \\ Furthermore, the present compound exhibits a comparatively large CW temperature together with an extended temperature regime of short-range spin correlations well above $T_{\rm N}=5.2$~K. It is worth noting that the emergence of two distinct regimes of short-range correlations, at $T_{\rm L}=10$ K and $T_{\rm H}=34$ K, suggests an underlying hierarchy of exchange interactions in the anisotropic hypertrillium lattice~\cite{PhysRevLett.125.167201}, although further investigations are needed to clarify this scenario. As a reference, a hierarchy of exchange couplings, $J_{4} > J_{5} > J_{3}$, has already been established in the isostructural compound KSrFe$_{2}$(PO$_{4}$)$_{3}$~\cite{k3lw-2567}. Within the mean-field approximation, the CW temperature estimated from the exchange constants of \(\mathrm{KSrFe_2(PO_4)_3}\) listed in Table~\ref{EXCH} is given by
		$\theta_{\mathrm{CW}}=\frac{S(S+1)}{3k_{\rm B}}\sum_i z_i J_i $.
		For \(z_1=1\) and \(z_2=z_3=z_4=z_5=3\), this yields effective \(\theta_{\mathrm{CW}} \approx -76~\mathrm{K}\) ~\cite{Gonzalez2024}.  Conversely, the larger $|\theta_{\mathrm{CW}}|$ observed in the present system, together with minimal disorder, is indicative of stronger exchange interactions in the underlying spin topology.  If the magnetism is interpreted in terms of a hypertrillium network with coupled dimers, the larger effective coordination would tend to increase the mean-field Curie--Weiss scale. Accordingly, using the exchange parameters from Table~\ref{EXCH} with $z_1=2$, $z_3=z_4=6$, and $z_5=3$, we obtain $\theta_{\mathrm{CW}} \approx -129~\mathrm{K}$, which may qualitatively align with the larger $|\theta_{\mathrm{CW}}|$ observed in the present compound. In any case, further theoretical and experimental investigations are required to clarify the origin of the strong exchange interactions. Taken together, these findings-highlighting spin frustration, short-range correlations, and persistent spin dynamics-along with the feasibility of high-quality single-crystal growth, provide a compelling basis for future inelastic neutron scattering experiments aimed at uncovering universal spin–spin correlation signatures in the hypertrillium lattice~\cite{PhysRevLett.125.167201}.
\section{CONCLUSION}

In summary, we have carried out a comprehensive investigation of the ground-state magnetic properties of 
K$_2$Fe$_2$(MoO$_4$)(PO$_4$)$_2$ single crystals using magnetization, specific heat, ESR, 
and $\mu$SR experiments. The combined experimental results provide clear evidence for strong antiferromagnetic correlations coexisting with weak magnetic order at $T_{\rm N}$ = 5.2 K. This scenario is corroborated by the $\lambda$-like anomaly in the magnetic susceptibility, $\mu$SR relaxation rate, and the wipeout of the ESR signal in the vicinity of $T_{\rm N}$.
 Above the ordering temperature, two distinct regimes 
of short-range spin correlations are identified at $T_{\rm H}=34$~K and $T_{\rm L}=10$~K, supported by anomalies in 
the magnetic specific heat and the temperature evolution of the ESR linewidth and $g$-factor. This observation alludes to a hierarchical exchange-energy landscape, resulting in the emergence of two distinct effective spin sublattices with different correlation scales. Furthermore, an 
additional anomaly at $T^{*}=3.2$~K suggests  the presence of  a reorientation of the ordered magnetic moments arising from weak subleading interactions that become operative below $T_{\rm N}$. Notably, $\mu$SR experiments confirm the persistence of dynamic spin fluctuations even below 
$T_{\rm N}$, underscoring the unconventional nature of the ordered state. The strong suppression of $T_{\rm N}$ under 
applied magnetic fields ($\mu_{0}H \geq 2$~T) raises the possibility of realizing a field-induced spin liquid state. Collectively, these results establish K$_2$Fe$_2$(MoO$_4$)(PO$_4$)$_2$ as a promising system for exploring the rich landscape of spin-liquid behavior and spin dynamics dictated by a hierarchical network of exchange interactions originating from three-dimensional frustrated couplings of coupled trillium lattices.

\section{Data Availability}

The data supporting the findings of this study are available upon reasonable request. However, the $\mu$SR data presented in this article
are openly accessible in Ref.~\cite{ISISdatabase}.

\section{ ACKNOWLEDGMENTS}
	The work at
SKKU was supported by the National Research Foundation (NRF) of Korea (Grants No. RS-2023-00209121 and
No. 2020R1A5A1016518).
 This work
was supported by HLD-HZDR, member of the European
Magnetic Field Laboratory (EMFL). R S acknowledges the financial support provided by the Ministry of Science and Technology in Taiwan under Project Numbers NSTC- 114-2124-M-001-009, NSTC-113-2112-M-001-045-MY3, Financial support from the Center of Atomic Initiative for New Materials (AI-Mat), National Taiwan University (Project No. 113L900801) and Academia Sinica for the budget of AS-iMATE-115-14. We thank the ISIS cryogenics team for their support throughout the beamtime, particularly for ensuring stable cryogenic performance during the low-temperature measurements.
\bibliography{Trillium.bib}
\end{document}